\documentstyle[amssymb,preprint,aps]{revtex}
\tightenlines

\begin{document}
\title{Isoscalar Hamiltonians for light atomic nuclei}
\author{G. P. Kamuntavi\v{c}ius$^{a,b,c}$, P. Navr\'{a}til$^{a,d}$, B. R. Barrett$%
^{a}$,}
\author{G. Sapragonaite$^{b}$, and R. K. Kalinauskas$^{c}$}
\address{$^{a}$Department of Physics, University of Arizona, Tucson, Arizona 85721\\
$^b$Vytautas Magnus University, Kaunas LT-3000, Lithuania\\
$^{c}$Institute of Physics, Vilnius LT-2600, Lithuania\\
$^{d}$Institute of Nuclear Physics, Academy of Sciences of the Czech\\
Republic,\\
250 68 \v{R}e\v{z} near Prague, Czech Republic}
\date{\today}
\maketitle

\begin{abstract}
The charge-dependent realistic nuclear Hamiltonian for a nucleus, composed
of neutrons and protons, can be successfully approximated by a
charge-independent one. The parameters of such a Hamiltonian, i.e., the
nucleon mass and the NN potential, depend upon the mass number A, charge Z
and isospin quantum number T of state of the studied nucleus.
\end{abstract}

\pacs{21.60.-n; 21.30.-x; 21.30.Fe.}

\section{INTRODUCTION}

The model of the atomic nucleus as a system composed of neutrons and protons
has been studied for over sixty years and has been found to provide a
more-or-less adequate description. By considering the constituents of the
nucleus as structureless identical fermions (i.e., nucleons), we obtain a
very useful formalism, which leads to significant simplifications of our
understanding and description of nuclear structure. These simplifications
follow from the fact that the neutron and the proton have similar masses and
that the potentials of the strong neutron-neutron (nn), neutron-proton (np)
and proton-proton (pp) interactions are also very similar. Over the
intervening years, simple models without correlations, such as the nuclear
shell model, have been unable to distinguish between different
parametrizations of the nucleon-nucleon (NN) interaction or to detect
details in the definition of the mass of the nucleon.

During the last decade, however, theoretical techniques for calculating
few-nucleon systems and the lightest atomic nuclei have advanced to the
point that the quality of the realistic NN potential is starting to play an
important role \cite{RMP98}. This necessarily involves the introduction of
charge dependence into the NN interaction. The most-recent realistic
potentials, such as the Nijmegen \cite{Nijm}, Argonne \cite{A-18} or CD-Bonn 
\cite{CDBonn}, are charge-dependent. On the one hand, the charge dependence
of the nuclear Hamiltonian complicates the description, but on the other,
the deviations of charge-dependent potentials from the charge-independent
ones, although non-negligible, are not very large. As a consequence, modern
calculations of light nuclei use as a starting point a Hamiltonian
preserving the isospin quantum number and include charge-dependent effects
perturbatively later on. The charge dependence of the kinetic and potential
energies of a real nucleus, however, can only be successfully taken into
account perturbatively, when the zeroth-order, charge-independent, isoscalar
nuclear Hamiltonian is as close as possible to the charge-dependent one. We
present here a method of construction for such a Hamiltonian.

In Sections II and III we show that the kinetic-energy operator of an atomic
nucleus composed of neutrons and protons can be precisely approximated by a
corresponding operator for a nucleus composed of nucleons, when the nucleon
mass is defined as the mean value of the quantum-mechanical operator for the
nucleon mass. The nucleon mass obtained in this way depends on the mass
number A and charge Z of the studied nucleus.

In Section IV we investigate the problem concerning the definition of the NN
interaction. For the sake of simplicity, we consider potentials, defined
separately in different two-nucleon channels, such as, the Reid'68 \cite
{Reid'68} and Nijmegen \cite{Nijm} potentials. The generalization of results
given by other realistic charge-dependent potentials, such as the CD-Bonn 
\cite{CDBonn} and Argonne-18 \cite{A-18} potentials, is straightforward. It
is shown that the best NN potential is a linear combination of the nn, np,
and pp potentials with coefficients, dependent not only on the nucleus
studied, i.e., $A$ and $Z$, but also on the isospin quantum number $T$ of
the state under consideration. So, the isoscalar Hamiltonians that we
introduce are different for different nuclei and even for different states
of the same nucleus.\ The derived values for the nucleon mass and the NN
interaction must satisfy minimal requirements (e.g., for systems composed
only of neutrons, the nucleon mass formula must give the neutron mass and
the NN interaction must reduce to the potential of the neutron-neutron
interaction) and must also be equal to the corresponding values for
non-trivial cases suggested for three-nucleon nuclei in earlier publications 
\cite{FriGib78}, \cite{Friar87}.

In Section V we present numerical tests of our recommendations and give our
conclusions in Section VI.

\section{THE MASS OF THE\ NUCLEON}

The nonrelativistic kinetic energy operator of the atomic nucleus equals 
\begin{equation}
K=-\frac{\hbar ^{2}}{2m_{p}}\sum_{i=1}^{Z}\nabla _{i}^{2}-\frac{\hbar ^{2}}{%
2m_{n}}\sum_{i=Z+1}^{A}\nabla _{i}^{2}+\frac{\hbar ^{2}}{2M}\nabla _{{\bf R}%
}^{2},  \label{1}
\end{equation}
where $m_{p}$ and $m_{n}$ are the proton and the neutron masses,
respectively; $M=Zm_{p}+Nm_{n}$\quad is the mass of the nucleus and 
\[
{\bf R}=\frac{1}{M}\left( \sum_{i=1}^{Z}m_{p}{\bf r}_{i}+%
\sum_{i=Z+1}^{A}m_{n}{\bf r}_{i}\right) 
\]
is the radius-vector of the center-of-mass of the nucleus .

The kinetic energy operator of the nucleus, as a system of nucleons is
simpler: 
\begin{equation}
K^{0}=-\frac{\hbar ^{2}}{2m}\sum_{i=1}^{A}\nabla _{i}^{2}+\frac{\hbar ^{2}}{%
2Am}\nabla _{{\bf R}_{0}}^{2},  \label{2}
\end{equation}
where $m$ is the nucleon mass and 
\[
{\bf R}_{0}=\frac{1}{A}\sum_{i=1}^{A}{\bf r}_{i} 
\]
is the center-of-mass radius-vector of the nucleus composed of nucleons.

Obviously, the operators $K$ and $K^{0}$ are different. In particular, the
expression for $K^{0}$ contains the parameter $m,$ which, in many
applications, is defined as an average of the proton and neutron masses;
i.e., 
\begin{equation}
\bar{m}=\frac{m_{n}+m_{p}}{2}.  \label{MM}
\end{equation}
On the other hand, calculations for a neutron and proton system involve a
transformation to relative and center-of-mass coordinates and a mass equal
to twice the reduced mass 
\begin{equation}
\mu =\frac{2m_{n}m_{p}}{m_{n}+m_{p}}.  \label{RM}
\end{equation}
While in the three nucleon case, it is sometimes assumed that $^{3}$H and $%
^{3}$He are composed of identical particles, each of mass $\frac{1}{3}\left(
2m_{n}+m_{p}\right) $ , $\frac{1}{3}\left( m_{n}+2m_{p}\right) $ \cite
{FriGib78}, respectively.

At first glance, it looks as though we can define the optimal kinetic energy
operator, which is as close as possible to a charge-dependent one, yet
charge-independent, by properly defining the mass of the nucleon in $K^{0}$.
However, the nucleon mass is not really a free parameter. It has to be
defined as the mean value of a nucleon mass operator: 
\[
\hat{m}=\frac{1}{A}\sum_{i=1}^{A}m\left( i\right) =\frac{1}{A}%
\sum_{i=1}^{A}\left( m_{n}P_{n}\left( i\right) +m_{p}P_{p}\left( i\right)
\right) , 
\]
where $m\left( i\right) $ is the mass of an $i$-th nucleon and 
\begin{equation}
P_{p}\left( i\right) =\frac{1}{2}-\tau _{z}\left( i\right) ,\text{\qquad }%
P_{n}\left( i\right) =\frac{1}{2}+\tau _{z}\left( i\right)  \label{Proj}
\end{equation}
are projectors upon proton and neutron states.

The mass of the nucleon in the nucleus $\left( A,Z\right) $ can be defined
exactly, because in the $A$-nucleon state, with a defined isospin
projection,\quad $M_{T}=\frac{1}{2}\left( N-Z\right) $ : 
\begin{eqnarray*}
m &=&\left\langle M_{T}\left| \hat{m}\right| M_{T}\right\rangle
=\left\langle M_{T}\left| \frac{1}{A}\sum_{i=1}^{A}m\left( i\right) \right|
M_{T}\right\rangle \\
&=&\left\langle M_{T}\left| \frac{m_{n}+m_{p}}{2}+\frac{\left(
m_{n}-m_{p}\right) }{A}\sum_{i=1}^{A}\tau _{z}\left( i\right) \right|
M_{T}\right\rangle \\
&=&\frac{m_{n}+m_{p}}{2}+\frac{\left( m_{n}-m_{p}\right) }{A}\left\langle
M_{T}\left| T_{z}\right| M_{T}\right\rangle \\
&=&\frac{m_{n}+m_{p}}{2}+\frac{\left( m_{n}-m_{p}\right) }{A}M_{T},
\end{eqnarray*}
or, in other words, 
\begin{equation}
m=\frac{1}{A}\left( Nm_{n}+Zm_{p}\right) .  \label{M1}
\end{equation}
This definition of the nucleon mass coincides with that used in the
above-mentioned three-nucleon case. For light nuclei with $N=Z$ , it equals
the mean value of the neutron and proton masses. For two nucleon systems it
coincides with twice the reduced mass in cases of two protons or two
neutrons and is very close to twice the neutron-proton reduced mass because 
\[
\mu -\bar{m}=\frac{2m_{n}m_{p}}{m_{n}+m_{p}}-\frac{m_{n}+m_{p}}{2}=-\bar{m}%
\delta ^{2}\approx -4.5\times 10^{-4}\frac{MeV}{c^{2}}, 
\]
where the widely used small parameter is: 
\begin{equation}
\delta =\frac{m_{n}-m_{p}}{m_{n}+m_{p}}\approx 6.887\,\times 10^{-4}.
\label{M2}
\end{equation}
Obviously, the difference between $\mu $ and $\bar{m}$ is negligible.
Realistic NN potentials, defined by using a reduced mass for the two
nucleons, cannot change significantly when the nucleon mass $m$, defined for
the two-nucleon system, is used instead of $\mu $.

\section{INTRINSIC VARIABLES AND\ OPERATORS}

A direct comparison of the operators $K^{0}$ and $K$ is impossible due to
different neutron and proton masses and different definitions of
center-of-mass operators. On the one hand, both operators, $K^{0}$ and $K,$
have redundant variables, because they are functions not only of one nucleon
but also of center-of-mass variables. On the other hand, both operators are
intrinsic; this means dependent only on intrinsic, translationally invariant
variables and independent of center-of-mass radius vectors. In order to
compare them, it is necessary to introduce intrinsic variables.

The Jacobi variables $\eta _{0},\eta _{1},...\eta _{A-1}$ for particles with
equal masses are well-known (see \cite{SShit} , \cite{FBS86} and references
therein). They can be defined by a corresponding Jacobi tree as an
orthonormal set. The expressions for $\eta _{\alpha }$ in terms of ${\bf r}%
_{i}$ and vice versa are as follows: 
\begin{equation}
{\large \eta }_{\alpha }=\sum_{i=1}^{A}a_{\alpha ,i}{\bf r}_{i},\hspace{1in}%
{\bf r}_{i}=\sum_{\alpha =0}^{A-1}{\large \eta }_{\alpha }a_{\alpha ,i}.
\label{3}
\end{equation}
Here ${\bf a}$ is an orthogonal $\left( A\times A\right) $ matrix : 
\begin{equation}
\sum_{i=1}^{A}a_{\alpha ,i}a_{\beta ,i}=\delta _{\alpha ,\beta },\hspace{1in}%
\sum_{\alpha =0}^{A-1}a_{\alpha ,i}a_{\alpha ,j}=\delta _{i,j}.  \label{3.1}
\end{equation}
The translational invariance of the intrinsic variables ${\bf \eta }_{1},...%
{\bf \eta }_{A-1}$ along with orthogonality gives the following conditions
for the matrix ${\bf a}$ : 
\begin{equation}
\sum_{i=1}^{A}a_{\alpha ,i}=\sqrt{A}\delta _{\alpha ,0},\qquad a_{0,i}=\frac{%
1}{\sqrt{A}}.  \label{3.2}
\end{equation}

Using these variables, we can rewrite the operator (\ref{2}) as follows: 
\begin{equation}
K^{0}=-\frac{\hbar ^{2}}{2m}\sum_{\alpha =1}^{A-1}\nabla _{{\bf \eta }%
_{\alpha }}^{2}.  \label{2.1}
\end{equation}

The analogous presentation for $K$ requires Jacobi variables for particles
with different proton and neutron masses. Let us introduce Jacobi variables
in the general case, in which the masses of all particles are different. The
reduction to protons and neutrons will then be straightforward. These
variables can be defined by the same kind of Jacobi tree, as in the previous
case.

The Jacobi tree by definition has $\left( 2A-1\right) $ vertices, $A$ of
which are vertices of the first degree (the degree of a vertex is defined as
the number of edges matching in this point). They should be arranged in a
line and marked with the one-particle radius-vectors ${\bf r}_{1},{\bf r}%
_{2},...,{\bf r}_{A}$ . In case of particles with different masses, however,
the vertices of the first degree of the tree have to be marked not only by
one-particle radius-vectors but also by corresponding masses $%
m_{1},m_{2},...,m_{A}$. The remaining $\left( A-1\right) $ vertices
(situated below the first group) determine the Jacobi coordinates ${\bf \xi }%
_{\alpha }$ $\left( \alpha =1,2,...,A-1\right) .$The degree of each equals
three, except for the deepest one which equals two. Contrary to the case of
equivalent particles, the orthogonal transformation can now be defined only
between Jacobi variables ${\bf \xi }_{0},{\bf \xi }_{1},...,{\bf \xi }_{A-1}$
and modified one-particle variables ${\bf x}_{1},{\bf x}_{2},...,{\bf x}%
_{A}, $ where ${\bf x}_{i}=\sqrt{m_{i}}{\bf r}_{i}.$ For the $\alpha $-th
vertex the Jacobi coordinate is 
\begin{equation}
{\bf \xi }_{\alpha }=\sqrt{\frac{m_{L}^{\left( \alpha \right) }m_{R}^{\left(
\alpha \right) }}{m_{L}^{\left( \alpha \right) }+m_{R}^{\left( \alpha
\right) }}}\left[ \frac{1}{m_{L}^{\left( \alpha \right) }}\sum_{i\in \left\{
L\right\} }\sqrt{m_{i}}{\bf x}_{i}-\frac{1}{m_{R}^{\left( \alpha \right) }}%
\sum_{i\in \left\{ R\right\} }\sqrt{m_{i}}{\bf x}_{i}\right] ,\qquad \left(
\alpha =1,2,...,A-1\right) .  \label{4}
\end{equation}
Here $\left\{ L\right\} $ is a manifold of the first-degree vertices, which
could be reached while moving from the $\alpha $-th vertex upwards along the
left edge, while $\left\{ R\right\} $ denotes the same for the right edge.
Obviously, the new Jacobi variables, ${\bf \xi }_{\alpha }$ $\left( \alpha
=1,2,...,A-1\right) $ , like the corresponding old ones, ${\bf \eta }%
_{\alpha }$ $\left( \alpha =1,2,...,A-1\right) ,$ are translationally
invariant, because they are defined as properly normalized radius vectors
between the centers-of-mass of two subtrees, defined by left and right edges
of the tree, as they are seen from the $\alpha $-th vertex. The masses of
left and right Jacobi clusters are defined in the following way: 
\[
m_{L}^{\left( \alpha \right) }=\sum_{i\in \left\{ L\right\} }m_{i},\hspace{%
1in}\ m_{R}^{\left( \alpha \right) }=\sum_{i\in \left\{ R\right\} }m_{i}. 
\]
This set has to be completed by a zero Jacobi variable, situated on the left
side of a second-degree vertex and proportional to a center-of-mass radius
vector: 
\begin{equation}
{\bf \xi }_{0}=\frac{1}{\sqrt{M}}\sum_{i=1}^{A}\sqrt{m_{i}}{\bf x}_{i},
\label{5}
\end{equation}
where $M=\sum_{i=1}^{A}m_{i}$ is the mass of the system. In general, 
\begin{equation}
{\bf \xi }_{\alpha }=\sum_{i=1}^{A}b_{\alpha ,i}\left(
m_{1},m_{2,}...,m_{A}\right) {\bf x}_{i},\hspace{1in}\left( \alpha
=0,1,...,A-1\right) ,  \label{7}
\end{equation}
where ${\bf b}$ is an orthogonal $\left( A\times A\right) $ matrix. In the
case of equivalent particles, this transformation coincides with that
defined in Eq. (\ref{3}) multiplied by $\sqrt{m}$. The proof of the
orthogonality of these transformations is given in Appendix A.

The tree sample for five particles is presented in Fig. 1. The matrix ${\bf b%
}$ for this tree equals\ 
\[
\left( 
\begin{array}{ccccc}
\sqrt{\frac{m_{1}}{M}} & \sqrt{\frac{m_{2}}{M}} & \sqrt{\frac{m_{3}}{M}} & 
\sqrt{\frac{m_{4}}{M}} & \sqrt{\frac{m_{5}}{M}} \\ 
\sqrt{\frac{m_{1}m_{45}}{Mm_{123}}} & \sqrt{\frac{m_{2}m_{45}}{Mm_{123}}} & 
\sqrt{\frac{m_{3}m_{45}}{Mm_{123}}} & -\sqrt{\frac{m_{4}m_{123}}{Mm_{45}}} & 
-\sqrt{\frac{m_{5}m_{123}}{Mm_{45}}} \\ 
\sqrt{\frac{m_{1}m_{3}}{m_{12}m_{123}}} & \sqrt{\frac{m_{2}m_{3}}{%
m_{12}m_{123}}} & -\sqrt{\frac{m_{12}}{m_{123}}} & 0 & 0 \\ 
\sqrt{\frac{m_{2}}{m_{12}}} & -\sqrt{\frac{m_{1}}{m_{12}}} & 0 & 0 & 0 \\ 
0 & 0 & 0 & \sqrt{\frac{m_{5}}{m_{45}}} & -\sqrt{\frac{m_{4}}{m_{45}}}
\end{array}
\right) , 
\]
where $m_{j...k}=\sum_{i=j}^{k}m_{i};\smallskip \ M=m_{12345}.$

Due to the orthogonality of the transformation from one-particle variables $%
{\bf x}_{1},{\bf x}_{2},...,{\bf x}_{A}$ to a system of Jacobi variables $%
{\bf \xi }_{0},{\bf \xi }_{1},...,{\bf \xi }_{A-1}$ , the transformation
from one set to another set of Jacobi variables, defined by different trees,
could be determined in the same way.

The kinetic energy operator (\ref{1}) in terms of these new variables can be
rewritten as 
\begin{equation}
K=-\frac{\hbar ^{2}}{2}\sum_{\alpha =1}^{A-1}\nabla _{{\bf \xi }_{\alpha
}}^{2}.  \label{1.1}
\end{equation}
A comparison of operators Eq.(\ref{2.1}) and Eq.(\ref{1.1}) requires the
transformation matrix between the sets ${\bf \xi }_{0},{\bf \xi }_{1},...,%
{\bf \xi }_{A-1}$ and ${\bf \eta }_{0},{\bf \eta }_{1},...{\bf \eta }_{A-1}$
. The result cannot depend on the choice of the Jacobi tree, so we can
choose the simplest one, given in Fig. 2, where protons are marked by the
numbers $1,...,Z$ and neutrons by the numbers $Z+1,...,Z+N=A$ . In such a
case 
\begin{eqnarray*}
{\bf \xi }_{\alpha } &=&\sum_{i=1}^{A}b_{\alpha ,i}\left(
m_{1},m_{2,}...,m_{A}\right) {\bf x}_{i}=\sum_{i=1}^{A}b_{\alpha ,i}\left(
m_{1},m_{2,}...,m_{A}\right) \sqrt{m_{i}}\sum_{\beta =0}^{A-1}a_{\beta ,i}%
{\bf \eta }_{\beta } \\
&=&\sum_{\beta =0}^{A-1}\left( \sum_{i=1}^{A}b_{\alpha ,i}\left(
m_{1},m_{2,}...,m_{A}\right) \sqrt{m_{i}}a_{\beta ,i}\right) {\bf \eta }%
_{\beta }.
\end{eqnarray*}
Due to the multipliers $\sqrt{m_{i}}$ , the matrix of the last
transformation is not an orthogonal matrix. This gives the following
expression for $K$ in terms of variables ${\bf \eta }_{\alpha }$: 
\[
\begin{array}{c}
K=-\frac{\hbar ^{2}}{2}\left[ \ \frac{ZN\left( m_{n}-m_{p}\right) ^{2}}{%
AMm_{n}m_{p}}\nabla _{{\bf \eta }_{0}}^{2}+\frac{2\sqrt{ZN}\left(
m_{n}-m_{p}\right) }{Am_{n}m_{p}}\left( \nabla _{{\bf \eta }_{0}}\cdot
\nabla _{{\bf \eta }_{1}}\right) \right. \\ 
\left. +\frac{M}{Am_{n}m_{p}}\nabla _{{\bf \eta }_{1}}^{2}+\frac{1}{m_{p}}%
\sum_{\alpha =2}^{Z}\nabla _{{\bf \eta }_{\alpha }}^{2}+\frac{1}{m_{n}}%
\sum_{\alpha =Z+1}^{A-1}\nabla _{{\bf \eta }_{\alpha }}^{2}\ \right] .
\end{array}
\]

So, the neutron and proton mass difference causes the intrinsic kinetic
energy operator to depend upon the center-of-mass variable. The
center-of-mass starts to move in an atomic nucleus composed of nucleons
instead of real protons and neutrons. However, this effect is not
significant, because the first term, proportional to $\nabla _{{\bf \eta }%
_{0}}^{2},$ has a very small coefficient, while the expectation value of the
second term, proportional to $\left( \nabla _{{\bf \eta }_{0}}\cdot \nabla _{%
{\bf \eta }_{1}}\right) ,$ vanishes either when the excitations of the
center-of-mass of a nucleus are under control or when the wave function is
intrinsic.

The difference in the kinetic-energy operators for neutron-proton and
nucleon nuclei is, thus, given by

\[
\begin{array}{c}
K-K^{0}=-\frac{\hbar ^{2}}{2m}\left[ \ \delta ^{2}\frac{ZN}{A}\frac{\left(
m_{n}+m_{p}\right) ^{2}}{Mm_{n}m_{p}}\nabla _{{\bf \eta }_{0}}^{2}+2\delta 
\frac{\sqrt{ZN}}{A}\frac{m_{n}+m_{p}}{m_{n}m_{p}}\left( \nabla _{{\bf \eta }%
_{0}}\cdot \nabla _{{\bf \eta }_{1}}\right) \right. \\ 
\left. +\left( \frac{Mm}{Am_{n}m_{p}}-1\right) \nabla _{{\bf \eta }%
_{1}}^{2}+\left( \frac{m}{m_{p}}-1\right) \sum_{\alpha =2}^{Z}\nabla _{{\bf %
\eta }_{\alpha }}^{2}+\left( \frac{m}{m_{n}}-1\right) \sum_{\alpha
=Z+1}^{A-1}\nabla _{{\bf \eta }_{\alpha }}^{2}\ \right] ,
\end{array}
\]
where $\delta $ is the parameter (\ref{M2}).

From the above expression, it is obvious that the mean value of the
difference in kinetic energies, estimated using intrinsic wave functions,
equals the sum of the mean values of the last three terms and can be
rewritten as: 
\begin{eqnarray}
\left\langle {\cal E}J\Pi M_{T}\left| K-K^{0}\right| {\cal E}J\Pi
M_{T}\right\rangle &=&\left[ \left( \frac{Mm}{Am_{n}m_{p}}-1\right) +\left( 
\frac{m}{m_{p}}-1\right) \left( Z-1\right) +\left( \frac{m}{m_{n}}-1\right)
\left( N-1\right) \right]  \nonumber \\
&&\times \left\langle {\cal E}J\Pi M_{T}\left| -\frac{\hbar ^{2}}{2m}\nabla
_{{\bf \eta }_{\alpha }}^{2}\right| {\cal E}J\Pi M_{T}\right\rangle ,
\label{6}
\end{eqnarray}
because the mean value of $\nabla _{{\bf \eta }_{\alpha }}^{2}$ is
independent of $\alpha $ (see Appendix B). The expression given in square
brackets equals 
\[
\frac{ZN\left( A-1\right) }{A^{2}}\frac{\left( m_{n}-m_{p}\right) ^{2}}{%
m_{n}m_{p}}\approx \frac{4ZN\left( A-1\right) }{A^{2}}\delta ^{2}, 
\]
while the matrix element has the same order as the kinetic energy of a
single nucleon in a light nucleus ( $\leq 40MeV$, see \cite{BorMot} ).
Consequently, 
\[
\left\langle K-K^{0}\right\rangle \lesssim 10^{-5}MeV. 
\]
This means that by using the defined value of the nucleon mass (\ref{M1}),
we can account for the n-p mass difference in the expression for the
intrinsic kinetic energy without any problems. In cases, when the mean value
of the neutron and proton masses (\ref{MM}) is used instead of the defined
value of the nucleon mass, this difference of expectation values (\ref{6})
is proportional to $\delta $ and can cause a noticeable difference in the
kinetic energies $K$ and $K^{0}$.

\section{THE POTENTIAL OF THE\ NUCLEON - NUCLEON INTERACTION}

The Hamiltonian of an atomic nucleus with a charge independence and charge
symmetry breaking NN potential has the form 
\[
H=K+V, 
\]
where $K$ is the translationally invariant operator of kinetic energy, Eq. (%
\ref{1}), and 
\[
V=\sum\limits_{i,k=1\left( i<k\right) }^{A}\left[ V_{pp}\left( i,k\right)
P_{p}\left( i\right) P_{p}\left( k\right) +V_{np}\left( i,k\right)
P_{n}\left( i\right) P_{p}\left( k\right) +V_{nn}\left( i,k\right)
P_{n}\left( i\right) P_{n}\left( k\right) \right] 
\]
is the potential-energy operator (the Coulomb interaction is included
in\quad $V_{pp}\left( i,k\right) $).

It is useful to employ the isospin formalism and the partial-waves expansion
in order to simplify the definition \ of the NN potential, i.e., 
\begin{eqnarray}
V\left( i,k\right) &\equiv &V\left( {\bf r}_{ik}\sigma _{i}\tau _{i}\sigma
_{k}\tau _{k}\right)  \nonumber \\
&=&\sum_{j\pi tm_{t}}V^{j\pi tm_{t}}\left( r_{ik}\right) P_{j\pi
tm_{t}}\left( \theta _{ik}\phi _{ik}\sigma _{i}\tau _{i}\sigma _{k}\tau
_{k}\right) ,  \label{14}
\end{eqnarray}
where\quad ${\bf r}_{ik}\equiv (r_{ik}\theta _{ik}\phi _{ik})={\bf r}_{i}-%
{\bf r}_{k}$ is the difference of radius-vectors of the i-th and the k-th
nucleons; $\sigma _{i}\tau _{i}$ are the spin and isospin variables of the $%
i $-th nucleon, respectively;\quad $j\pi t$ are the quantum numbers - total
momentum, parity and isospin - of the two nucleon state; and $P_{j\pi
tm_{t}}\left( \theta _{ik}\phi _{ik}\sigma _{i}\tau _{i}\sigma _{k}\tau
_{k}\right) $ is the projection operator on this state. The correspondence
between our notation,\quad $j\pi t,$ and conventional spectroscopic
identifiers\quad $^{2s+1}L_{j}$ is: 
\[
0^{+}1\sim ^{1}S_{0},\text{\quad }0^{-}1\sim ^{3}P_{0,}\text{\quad }%
1^{+}0\sim ^{3}S_{1}-^{3}D_{1},\text{\quad }1^{-}1\sim ^{3}P_{1},\quad
1^{-}0\sim ^{1}P_{1},\ldots . 
\]
Also, $m_{t}$ stands for the two-nucleon isospin projection;\quad $m_{t}=1$
corresponds to two neutrons;\quad $m_{t}=0$, to neutron-proton; and\quad $%
m_{t}=-1$, to two protons. Traditionally, NN potentials are constructed by
fitting np ($m_{t}=0$) data for\quad $t=0$ states and either np ($m_{t}=0$)
or pp ($m_{t}=-1$) data for\quad $t=1$ states \cite{BorMot}.

In general, the realistic potentials can be different for\quad $m_{t}=0,\pm
1 $ in each two-nucleon channel with\quad $t=1$. To some extent it holds
even for charge-symmetric potentials of the strong interaction, because the
Coulomb interaction must to be included in\quad $V^{j\pi
(t=1)(m_{t}=-1)}\left( r_{ik}\right) $. In channels with $\pi =\left(
-1\right) ^{j}$ the potential $V^{j\pi tm_{t}}\left( r_{ik}\right) $ is one
function of $r_{ik}$ , when $t=0,$ and three different functions of $r_{ik}$%
, when $t=1$ (the best example being $^{1}$S$_{0}$)$.$ In channels with $\pi
=\left( -1\right) ^{j+1}$ and $t=0,$ the potential $V^{j\pi tm_{t}}\left(
r_{ik}\right) $ can be given as three different functions of $r_{ik}$ ,
namely entries in a second-order symmetric matrix (such as the
parametrization in the channel $1^{+}0\sim ^{3}S_{1}-^{3}D_{1}$ , as given
for the Reid68 \cite{Reid'68} and Nijmegen \cite{Nijm} potentials) or nine
different functions of $r_{ik}$, corresponding to different values of $m_{t}$
when $\pi =\left( -1\right) ^{j+1}$ and $t=1$. The first channel of this
kind is $2^{-}1.$ This parametrization, however, is performed as three
different functions of $r_{ik}$. As mentioned earlier, they obviously are
different for $m_{t}=-1$ in comparison to $m_{t}=0$ or $m_{t}=+1,$ due to
Coulomb interaction of protons.

The charge-independent Hamiltonian of an atomic nucleus composed of nucleons
is 
\begin{equation}
H^{0}=K^{0}+V^{0}  \label{H0}
\end{equation}
with the kinetic energy operator, defined in Eq.(\ref{2}), and the
charge-independent potential-energy operator 
\begin{equation}
V^{0}=\sum\limits_{i,k=1\left( i<k\right) }^{A}V^{0}\left( i,k\right) ,
\label{V0}
\end{equation}
whose partial-wave expansion is given by 
\begin{eqnarray}
V^{0}\left( i,k\right) &\equiv &V^{0}\left( {\bf r}_{ik}\sigma _{i}\tau
_{i}\sigma _{k}\tau _{k}\right)  \nonumber \\
&=&\sum_{j\pi t}V^{j\pi t}\left( r_{ik}\right) P_{j\pi t}\left( \theta
_{ik}\phi _{ik}\sigma _{i}\tau _{i}\sigma _{k}\tau _{k}\right) .
\label{14.1}
\end{eqnarray}

As was shown above, there are no problems with regard to the
charge-dependence of the kinetic energy, because by taking the proper value
of the nucleon mass, one can use $K^{0}$ instead of\quad $K$ in the
expression for\quad $H,$ due to the negligible difference between these two
kinetic energy operators$.$ In such a case, the Hamiltonian\quad $H$ can be
expressed in the form 
\begin{equation}
H=\sum\limits_{i,k=1\left( i<k\right) }^{A}h\left( i,k\right) ,  \label{H1}
\end{equation}
where 
\begin{equation}
h\left( i,k\right) =-\frac{\hbar ^{2}}{2mA}\left( \nabla _{i}-\nabla
_{k}\right) ^{2}+V\left( {\bf r}_{i}-{\bf r}_{k},\sigma _{i}\tau _{i}\sigma
_{k}\tau _{k}\right) .  \label{RHP}
\end{equation}

The {\em exact} expression for an arbitrary eigenvalue of such an operator
is ( see \cite{Kam81}, \cite{Kam89} and references therein) : 
\begin{equation}
{\cal E}^{J\Pi M_{T}}=\left\langle H\right\rangle _{{\cal E},{\cal E}}^{J\Pi
M_{T}}=\frac{A\left( A-1\right) }{2}\sum_{\varepsilon ,j\pi
tm_{t}}\varepsilon ^{j\pi tm_{t}}Q_{\varepsilon ,j\pi tm_{t}}\left( {\cal E}%
J\Pi M_{T}\right) ,  \label{15}
\end{equation}
where the sum runs over all the two nucleon states\quad $j\pi t$ and all
eigenvalues\quad $\varepsilon $ of the Reduced Hamiltonian (RH) operator 
\begin{equation}
h\equiv h\left( A-1,A\right) =-\frac{2\hbar ^{2}}{mA}\Delta _{{\bf r}%
}+V\left( {\bf r},\sigma _{A-1}\tau _{A-1}\sigma _{A}\tau _{A}\right)
\label{16}
\end{equation}
with ${\bf r=r}_{A-1}-{\bf r}_{A}$.

The $Q_{\varepsilon ,j\pi tm_{t}}\left( {\cal E}J\Pi M_{T}\right) $ are
diagonal entries of the intrinsic density matrix in terms of the RH
eigenfunctions: 
\begin{equation}
Q_{\varepsilon ,j\pi tm_{t}}\left( {\cal E}J\Pi M_{T}\right) \equiv
Q_{\varepsilon ,j\pi tm_{t},\varepsilon ,j\pi tm_{t}}\left( {\cal E}J\Pi
M_{T}\right) ,  \label{tmm}
\end{equation}
\begin{eqnarray}
Q_{\varepsilon ,j\pi tm_{t},\varepsilon ^{\prime },j^{\prime }\pi ^{\prime
}t^{\prime }m_{t}^{\prime }}\left( {\cal E}J\Pi M_{T}\right) &=&\sum_{\sigma
_{1}\tau _{1}...\sigma _{A-1}\tau _{A-1}\sigma _{A}\tau _{A}\sigma
_{A-1}^{\prime }\tau _{A-1}^{\prime }\sigma _{A}^{\prime }\tau _{A}^{\prime
}}\int d{\bf \xi }_{1}...d{\bf \xi }_{A-2}d{\bf \xi }_{A-1}d{\bf \xi }%
_{A-1}^{\prime }  \nonumber \\
&&\times \psi _{_{\varepsilon ,j\pi tm_{t}}}^{+}\left( {\bf \xi }%
_{A-1}\sigma _{A-1}\tau _{A-1}\sigma _{A}\tau _{A}\right)  \nonumber \\
&&\times \Psi _{{\cal E}J\Pi M_{T}}\left( {\bf \xi }_{1}...{\bf \xi }_{A-2}%
{\bf \xi }_{A-1}\sigma _{1}\tau _{1}...\sigma _{A-1}\tau _{A-1}\sigma
_{A}\tau _{A}\right)  \nonumber \\
&&\times \Psi _{{\cal E}J\Pi M_{T}}^{+}\left( {\bf \xi }_{1}...{\bf \xi }%
_{A-2}{\bf \xi }_{A-1}^{\prime }\sigma _{1}\tau _{1}...\sigma _{A-1}^{\prime
}\tau _{A-1}^{\prime }\sigma _{A}^{\prime }\tau _{A}^{\prime }\right) 
\nonumber \\
&&\times \psi _{_{\varepsilon ^{\prime },j^{\prime }\pi ^{\prime }t^{\prime
}m_{t}^{\prime }}}\left( {\bf \xi }_{A-1}^{\prime }\sigma _{A-1}^{\prime
}\tau _{A-1}^{\prime }\sigma _{A}^{\prime }\tau _{A}^{\prime }\right) ,
\label{tmd}
\end{eqnarray}
where $\Psi _{{\cal E}J\Pi M_{T}}\left( {\bf \xi }_{1}...{\bf \xi }%
_{A-1}\sigma _{1}\tau _{1}...\sigma _{A}\tau _{A}\right) $ is the intrinsic
wave-function of the nucleus.

According to the definition of the density matrix 
\begin{equation}
Q_{\varepsilon ,j\pi tm_{t}}\left( {\cal E}J\Pi M_{T}\right) \geq 0,\quad 
\text{and}\quad \sum_{\varepsilon ,j\pi tm_{t}}Q_{\varepsilon ,j\pi
tm_{t}}\left( {\cal E}J\Pi M_{T}\right) =1.  \label{117}
\end{equation}

These are probabilities of definite states of two nucleon relative
motion,\quad $\varepsilon j\pi tm_{t},$ in the state under
investigation,\quad ${\cal E}J\Pi M_{T},$ i.e., the state specified by the
exact quantum numbers - energy, total momentum, parity and isospin
projection of the nucleus $\left( A,Z\right) $ .

The expression for the eigenvalue of the Hamiltonian 
\begin{equation}
H^{0}=\sum\limits_{i,k=1\left( i<k\right) }^{A}h^{0}\left( i,k\right) ,
\label{H00}
\end{equation}
where 
\begin{equation}
h^{0}\left( i,k\right) =-\frac{\hbar ^{2}}{2mA}\left( \nabla _{i}-\nabla
_{k}\right) ^{2}+V^{0}\left( {\bf r}_{i}-{\bf r}_{k},\sigma _{i}\tau
_{i}\sigma _{k}\tau _{k}\right)  \label{RH0}
\end{equation}
is very similar to that given by Eq. (\ref{15}): 
\begin{equation}
{\cal E}^{J\Pi T}=\left\langle H^{0}\right\rangle _{{\cal E},{\cal E}}^{J\Pi
T}=\frac{A\left( A-1\right) }{2}\sum_{\varepsilon ,j\pi t}\varepsilon ^{j\pi
t}Q_{\varepsilon ,j\pi t}\left( {\cal E}J\Pi T\right) ,  \label{18}
\end{equation}
but is slightly simpler, because the corresponding NN\ potential is
charge-independent. As a consequence, the total isospin $T$ is an exact
quantum number. Moreover, in such a case, the eigenvalues are the same for
all nuclei of the isospin multiplet and, hence, independent on $M_{T}$ .
Here 
\begin{equation}
h^{0}\equiv h^{0}\left( A-1,A\right) =-\frac{2\hbar ^{2}}{mA}\Delta _{{\bf r}%
}+V^{0}\left( {\bf r},\sigma _{A-1}\tau _{A-1}\sigma _{A}\tau _{A}\right) .
\label{19}
\end{equation}

A straightforward comparison of the eigenvalues, Eqs. (\ref{15}) and (\ref
{18}), is impossible, as eigenfunctions of the Hamiltonians,\quad $H$
and\quad $H^{0},$ do not coincide. Eigenvalues and eigenfunctions of the
corresponding Reduced Hamiltonians (Eqs. (\ref{16}) and (\ref{19})) are also
different. To compare them, it is necessary to define kind of equivalent
potential $V^{0}$, for which we are looking. From partial-wave
decompositions of the potentials, Eqs. (\ref{14}) and (\ref{14.1}), it
follows, that the realistic potential in the states with\quad $t=0$ is the
same in both cases, because\quad $m_{t}$ can take only the value zero.
If\quad $t=1$, there are three different pairs of nucleons ($m_{t}=0,\pm 1$)
and the NN potential\quad $V^{j\pi tm_{t}}\left( r_{ik}\right) $ is
different in each case. Therefore, the symmetric, charge-independent
potential in an arbitrary two-nucleon state can be defined as the sum of
these charge-dependent potentials: 
\begin{equation}
V^{j\pi t}\left( r_{ik}\right) =\sum_{m_{t}}c_{tm_{t}}V^{j\pi tm_{t}}\left(
r_{ik}\right) ,  \label{20}
\end{equation}
where the $c_{tm_{t}}$ are normalized coefficients: 
\begin{equation}
\sum_{m_{t}}c_{tm_{t}}=1.  \label{20.0}
\end{equation}

The optimal values of these coefficients are (see Appendix C): 
\begin{eqnarray}
c_{00} &=&1,  \label{25} \\
c_{1-1} &=&\frac{\left( A-2M_{T}\right) \left( A-2M_{T}-2\right) }{3A\left(
A-2\right) +4T\left( T+1\right) }=\frac{4Z\left( Z-1\right) }{3A\left(
A-2\right) +4T\left( T+1\right) },  \label{26} \\
c_{10} &=&\frac{A\left( A-2\right) +4T\left( T+1\right) -8M_{T}^{2}}{%
3A\left( A-2\right) +4T\left( T+1\right) }=\frac{8NZ-A\left( A+2\right)
+4T\left( T+1\right) }{3A\left( A-2\right) +4T\left( T+1\right) },
\label{27} \\
c_{1+1} &=&\frac{\left( A+2M_{T}\right) \left( A+2M_{T}-2\right) }{3A\left(
A-2\right) +4T\left( T+1\right) }=\frac{4N\left( N-1\right) }{3A\left(
A-2\right) +4T\left( T+1\right) }.  \label{28}
\end{eqnarray}

These values, obtained as a result of consideration of probability
distributions, have a rather simple interpretation. Having in mind the
definition of marginal probabilities, as given in \cite{Kam81}, and the
definition of the $c_{tm_{t}}$ , as given in (C5), we find that these
coefficients equal the number of pairs of nucleons with the isospin and
corresponding projection quantum numbers, $t$ and $m_{t}$, divided by the
number of pairs with the isospin quantum number $t$.

Finally, the charge independent Hamiltonian can be expressed in the form 
\begin{equation}
H^{0}\left( A,T,M_{T}\right) =\sum_{i,k=1(i<k)}^{A}h^{0}\left(
A,T,M_{T};i,k\right) ,  \label{29}
\end{equation}
where 
\begin{equation}
h^{0}\left( A,T,M_{T};i,k\right) =-\frac{\hbar ^{2}}{2m\left( A,M_{T}\right)
A}\left( \nabla _{i}-\nabla _{k}\right) ^{2}+V^{0}\left( A,T,M_{T};{\bf r}%
_{i}-{\bf r}_{k},\sigma _{i}\tau _{i}\sigma _{k}\tau _{k}\right) .
\label{30}
\end{equation}

The eigenvalue of $H^{0}\left( A,T,M_{T}\right) $ is very similar to that
given by Eq. (\ref{18}), but now has the form 
\begin{equation}
{\cal E}^{J\Pi T}\left( A,T,M_{T}\right) =\left\langle H^{0}\left(
A,T,M_{T}\right) \right\rangle _{{\cal E},{\cal E}}^{J\Pi T}=\frac{A\left(
A-1\right) }{2}\sum_{\varepsilon ,j\pi t}\varepsilon ^{j\pi t}\left(
A,T,M_{T}\right) Q_{\varepsilon ,j\pi t}\left( {\cal E}J\Pi T\right) .
\label{31}
\end{equation}

From the last expression we observe that our result for a charge-independent
Hamiltonian, which includes charge dependent effects (e.g., Coulomb
interaction, charge dependence of the potential and the neutron and proton
mass difference), is a function of the parameters $A,T,$ and $M_{T}.$
Comparing with the exact expression (\ref{15}), we observe that isospin now
appears as an exact quantum number and the density matrix is independent of $%
M_{T}$. Consequently, the density matrix is the same for all members of an
isospin multiplet, as it must be for a charge independent Hamiltonian.
Contrary to the commonly used expression for charge-independent potentials,
i.e., Eq. (\ref{18}), the present effective Hamiltonians (Eq. (\ref{29}))
are different for different nuclei of an isospin multiplet and for states of
the same nucleus with different values of total isospin $T.$

\section{NUMERICAL TESTS}

The effective Hamiltonians (Eq. (\ref{29})), obtained above, agree with the
well-known, trivial isoscalar Hamiltonians for nuclear matter and for states
of finite nuclei with total isospin\quad $T=0$ , because, in this case, the
nucleon mass equals half of neutron and proton masses and the effective
potential is the average of the three potentials. However, even in this
case, the recommendations given above allow us to express in isoscalar form
the Coulomb interaction of the protons. In the light-nuclei region for
isospin multiplets with\quad $T\neq 0,$ the values of the nucleon mass and
the effective potential are different for nuclei with different\quad $M_{T}$
values.

For the triton ($A=3,$ $T=1/2,$ $M_{T}=1/2$), for example, the mass of the
nucleon is 
\[
m=\frac{2}{3}m_{n}+\frac{1}{3}m_{p}\;, 
\]
and the equivalent charge-independent potential is 
\[
V^{j\pi \left( t=1\right) }\left( r\right) =\frac{1}{3}V^{j\pi
(t=1)(m_{t}=0)}\left( r\right) +\frac{2}{3}V^{j\pi (t=1)(m_{t}=1)}\left(
r\right) \;, 
\]
or simply 
\[
V\left( r\right) =\frac{1}{3}V_{np}^{t=1}\left( r\right) +\frac{2}{3}%
V_{nn}\left( r\right) \;, 
\]
which agrees with the results obtained in \cite{FriGib78} using the same
nucleon mass and in \cite{Friar87} for the charge-independent potential in
the\quad $^{1}S_{0}$ state.

For $^{3}$He ($A=3,$ $T=1/2,$ $M_{T}=-1/2$) the mass of nucleon equals 
\[
m=\frac{1}{3}m_{n}+\frac{2}{3}m_{p}\;. 
\]
The equivalent charge-independent potential is 
\[
V^{j\pi \left( t=1\right) }\left( r\right) =\frac{1}{3}V^{j\pi
(t=1)(m_{t}=0)}\left( r\right) +\frac{2}{3}V^{j\pi (t=1)(m_{t}=-1)}\left(
r\right) \;, 
\]
with the Coulomb interaction included in the last term.

In order to test the results obtained in section IV, we performed several
calculations in the framework of the large-basis, no-core shell-model
approach \cite{NB96}, using interactions that directly break isospin as well
as the corresponding NN interactions given by formulas (\ref{25})-(\ref{28}%
). The effective interactions were derived from the free np, pp, and nn
interactions. These effective interactions then served as input for the
large-basis shell-model calculations of the $A=3$ system, e.g., $^{3}$H and $%
^{3}$He, and of the $A=6$ system, e.g., $^{6}$Li, $^{6}$He, $^{6}$Be. In the
present calculations we employed the CD-Bonn \cite{CDBonn} potential, which
includes isospin symmetry breaking terms. In addition, the Coulomb potential
was added to the proton-proton interaction. To get energy convergence, one
needs $N_{{\rm \max }}>30$ for the three-nucleon problem. However, for the
purpose of the present work it is enough to perform the shell-model
calculations in a sufficiently large, but still easily accessible model
space. We chose $N_{{\rm \max }}=8$, which corresponds to $8\hbar \Omega $
excitations above the unperturbed ground state for the $A=3$ system and $%
6\hbar \Omega $ excitations for the $A=6$ system. For the
harmonic-oscillator frequency, we picked $\hbar \Omega =19$ MeV for the $A=3$
system and $\hbar \Omega =17.2$ MeV for the $A=6$ system. These choices
follow from the phenomenological formula $\hbar \Omega =45A^{-\frac{1}{3}%
}-25A^{-\frac{2}{3}}$ MeV.

The wave-functions obtained in the no-core shell-model approach satisfy all
the requirements necessary for realistic nuclear wave functions, i.e., the
excitation of the center-of-mass of the nucleus is under control and they
correspond to exact eigenvalues with good quantum numbers $J,\Pi ,$ (and $T$
in the case of the isospin invariant effective interaction). The results of
our shell-model calculations are presented in Table\ I. These results
demonstrate that, in general, the suggested definitions for the nucleon mass
and the NN interactions are successful. The deviations between the exact
(with charge-symmetry breaking interactions) and the approximate (with
isospin invariant interactions) calculations are at the level of a fraction
of a percent. The best correspondence between the results for the
charge-dependent and charge-independent Hamiltonians is obtained for $^{3}$H
and $^{3}$He and for all six states of $^{6}$Li. Moreover, we performed
calculations for the T=1 states in $^{6}$Li, using the isoscalar potential
for T=0. Analogous calculations for the T=0 states of the same nucleus were
also performed utilizing the isoscalar potential for T=1. In all cases the
absolute values of differences between charge-dependent and
charge-independent results are some three times larger in comparison with
calculations using the correct value of the isospin for the state under
investigation. As can be seen from the results in Table I, the biggest
difference between energies is obtained for $^{6}$Be, where it equals 1-2\%
of the ''exact'' value, and, to some extent, in $^{6}$He, where it does not
exceed 1\%.

\section{SUMMARY\ AND\ CONCLUSIONS}

As is well-known, charge symmetry in nuclei is broken at the very least due
to the difference between the proton and neutron masses, the
charge-dependence of the strong interaction and the Coulomb interaction
among the protons. These effects are not very large for light nuclei in
comparison with the binding energy of the entire nucleus, but there are
cases, in which they are significant and play a very important role, such as
for nuclei near drip-lines, which are loosely bound. However, the
description of nuclei using a fully charge-dependent Hamiltonian is
complicated, so the construction of a Hamiltonian, as close as possible to
charge-independent one, would be useful.

The present article is devoted to the consideration of this problem. It is
shown that charge-dependence of the realistic Hamiltonian for light nuclei
can be taken into account by replacing the realistic one by an equivalent
charge-independent Hamiltonian. The parameters of such a Hamiltonian (i.e.,
the nucleon mass and the NN potential) are nucleus and state dependent. Our
results can be formulated as simple recommendations for the construction of
such an isoscalar Hamiltonian:

\begin{enumerate}
\item  Take the nucleon mass equal to 
\[
m=\frac{1}{A}\left( Nm_{n}+Zm_{p}\right) . 
\]

\item  Take the realistic NN potential in states with two-nucleon isospin $%
t=0$ as they are defined from the neutron-proton data analyze.

\item  Take the isoscalar NN potential in states with $t=1$ as a linear
combination, Eq. (\ref{20}), of the neutron-neutron $\left( m_{t}=1\right) $%
, neutron-proton $\left( m_{t}=0\right) $, and proton-proton (with Coulomb
potential included) $\left( m_{t}=-1\right) $ realistic potentials with
coefficients $c_{tm_{t}}$, given in Eqs. (\ref{25})-(\ref{28}).
\end{enumerate}

The eigenfunctions of such a Hamiltonian have the isospin $T$ of the nucleus
as good a quantum number, but are different for the individual members of
the isospin multiplet due to the dependence of the isoscalar Hamiltonians on 
$M_{T}$.

In general, the results obtained are self-consistent and do not conflict
with the simple requirements mentioned in the Introduction. Moreover, the
isoscalar part of the Coulomb interaction has a nicer form than the earlier
result suggested in \cite{PPCW97}, Eq. (2.21), because the isoscalar Coulomb
interaction of the nucleons in the present work equals zero in two-nucleon
states with isospin $t=0$. Consequently, the normalization of this
interaction to the number of proton pairs occurs naturally.

Our results can be considered to be quite good in all cases considered,
because, due to our normalization conditions, Eq. (\ref{20.0}), we really
have only two free parameters for the construction of the isoscalar
interaction. The approximation given here is a good starting point for a
perturbational account of nuclear charge-dependent effects not included in
the isoscalar Hamiltonian.

\section{ACKNOWLEDGMENTS}

This work was supported by NSF grants No. PHY96-05192 and INT98-06614. P.N.
also acknowledges partial support from a grant of the Grant Agency of the
Czech Republic 202/96/1562. G.P.K. acknowledges the CIES for a Fulbright
Research Fellowship while at the University of Arizona and partial support
from a grant No 391 of the Lithuanian State Science and Studies Foundation.
G.P.K. and R.K.K. would like to thank Bruce R. Barrett and the University of
Arizona for their hospitality.

\section{APPENDIX A}

The orthogonality of the transformation Eq. (\ref{7}) follows directly from
the observation that an arbitrary Jacobi tree can be written as a system of
two subtrees, Fig. 3. The first (left) subtree with first-degree vertices $%
{\bf r}_{1}m_{1},$ ${\bf r}_{2}m_{2},...,{\bf r}_{q}m_{q}$ defines the
Jacobi variables ${\bf \xi }_{2},{\bf \xi }_{3},...,{\bf \xi }_{q}$ , the
second (right) subtree contains ${\bf r}_{q+1}m_{q+1},$ ${\bf r}%
_{q+2}m_{q+2},...,{\bf r}_{q+t}m_{q+t}$ and ${\bf \xi }_{q+1},{\bf \xi }%
_{q+2},...,{\bf \xi }_{q+t-1}$ , correspondingly. Obviously, it is possible
to introduce in the first and second cases the additional Jacobi variables
proportional to the center-of-mass coordinates of the subtrees, defined as 
\[
{\bf \xi }_{-1}=\frac{1}{\sqrt{m_{1...q}}}\sum_{i=1}^{q}\sqrt{m_{i}}{\bf x}%
_{i};\quad \text{and}\quad {\bf \xi }_{-q}=\frac{1}{\sqrt{m_{q+1...q+t}}}%
\sum_{i=q+1}^{q+t}\sqrt{m_{i}}{\bf x}_{i}. 
\]
The indices of these coordinates are negative, according to the rule that
the arbitrary center-of-mass Jacobi variable situated near an intrinsic
Jacobi variable ${\bf \xi }_{\alpha }$ is labelled as ${\bf \xi }_{1-\alpha
} $ (${\bf \xi }_{1}$ and ${\bf \xi }_{0}$, ${\bf \xi }_{2}$ and ${\bf \xi }%
_{-1}$ , ${\bf \xi }_{q+1}$ and ${\bf \xi }_{-q}$ ). Suppose, the
transformations from ${\bf r}_{1}m_{1},$ ${\bf r}_{2}m_{2},...,{\bf r}%
_{q}m_{q}$ to ${\bf \xi }_{-1},{\bf \xi }_{2},{\bf \xi }_{3},...,{\bf \xi }%
_{q}$ and from ${\bf r}_{q+1}m_{q+1},$ ${\bf r}_{q+2}m_{q+2},...,{\bf r}%
_{q+t}m_{q+t}$ to ${\bf \xi }_{-q},{\bf \xi }_{q+1},{\bf \xi }_{q+2},...,%
{\bf \xi }_{q+t-1}$ are orthogonal. Then the complete transformation is also
orthogonal, because the transformation matrix from ${\bf \xi }_{-1},{\bf \xi 
}_{-q}$ to ${\bf \xi }_{0},{\bf \xi }_{1}$ is orthogonal: 
\[
\left( 
\begin{array}{c}
{\bf \xi }_{0} \\ 
{\bf \xi }_{1}
\end{array}
\right) =\left( 
\begin{array}{cc}
\sqrt{\frac{m_{1..q}}{M}} & \sqrt{\frac{m_{q+1...q+t}}{M}} \\ 
\sqrt{\frac{m_{q+1...q+t}}{M}} & -\sqrt{\frac{m_{1...q}}{M}}
\end{array}
\right) \left( 
\begin{array}{c}
{\bf \xi }_{-1} \\ 
{\bf \xi }_{-q}
\end{array}
\right) . 
\]

Following this process to the top of the tree, we can arrive at orthogonal
two-dimensional transformations. Therefore, the complete transformation $%
\left( q+t\right) \times \left( q+t\right) $ is orthogonal.

\section{APPENDIX B}

The independence of the matrix element, Eq. (\ref{6}), $\left\langle {\cal E}%
J\Pi M_{T}\left| \nabla _{{\bf \eta }_{\alpha }}^{2}\right| {\cal E}J\Pi
M_{T}\right\rangle $, of $\alpha $, the intrinsic Jacobi variable number $%
\left( \alpha =1,2,...,A-1\right) ,$ can be shown as follows. By definition,
this element equals the product of sums over spin and isospin variables and
of integrals over intrinsic Jacobi variables 
\begin{eqnarray*}
&&\sum_{\sigma _{1,}\sigma _{2}...\sigma _{A}}\sum_{\tau _{1},\tau
_{2}...\tau _{A}}\int d{\bf \eta }_{1}\int d{\bf \eta }_{2}...\int d{\bf %
\eta }_{A-1}\Psi _{{\cal E}J\Pi M_{T}}^{\ast }\left( {\bf \eta }_{1}{\bf %
\eta }_{2}...{\bf \eta }_{A-1}\sigma _{1,}\sigma _{2}...\sigma _{A}\tau
_{1},\tau _{2}...\tau _{A}\right) \\
&&\times \nabla _{{\bf \eta }_{\alpha }}^{2}\Psi _{{\cal E}J\Pi M_{T}}\left( 
{\bf \eta }_{1}{\bf \eta }_{2}...{\bf \eta }_{A-1}\sigma _{1,}\sigma
_{2}...\sigma _{A}\tau _{1},\tau _{2}...\tau _{A}\right) ,
\end{eqnarray*}
where $\Psi _{{\cal E}J\Pi M_{T}}\left( ...\right) $ is an intrinsic wave
function of an atomic nucleus. This wave function is antisymmetric with
respect to all nucleons. However, this property in Jacobi variables is not
expandable in a simple way, because permutations of one-nucleon variables $%
{\bf r}_{i}$ generate orthogonal transformations of Jacobi variables. We can
simplify by multiplying the given expression by the integral 
\[
\int d{\bf \eta }_{0}\Theta ^{\ast }\left( {\bf \eta }_{0}\right) \Theta
\left( {\bf \eta }_{0}\right) \equiv 1, 
\]
where $\Theta \left( {\bf \eta }_{0}\right) $ is an arbitrary, normalized
wave function of the center-of-mass. The new wave functions 
\[
\Theta \left( {\bf \eta }_{0}\right) \Psi _{{\cal E}J\Pi M_{T}}\left( {\bf %
\eta }_{1}{\bf \eta }_{2}...{\bf \eta }_{A-1}\sigma _{1,}\sigma
_{2}...\sigma _{A}\tau _{1},\tau _{2}...\tau _{A}\right) 
\]
can be rewritten as usual antisymmetric functions of one-particle variables 
\[
\Phi _{{\cal E}J\Pi M_{T}}\left( {\bf r}_{1}{\bf r}_{2}...{\bf r}_{A}\sigma
_{1,}\sigma _{2}...\sigma _{A}\tau _{1},\tau _{2}...\tau _{A}\right) . 
\]
Due to the orthogonality of the matrix ${\bf a}$ , Eq. (\ref{3}), the
integrals can be transformed into one-particle variables: 
\[
\int d{\bf \eta }_{0}\int d{\bf \eta }_{1}\int d{\bf \eta }_{2}...\int d{\bf %
\eta }_{A-1}=\int d{\bf r}_{1}\int d{\bf r}_{2}...\int d{\bf r}_{A}. 
\]

The operator can be written in the form 
\[
\nabla _{{\bf \eta }_{\alpha }}^{2}=\sum_{i,j=1}^{A}a_{\alpha ,i}a_{\alpha
,j}\left( \nabla _{i}\cdot \nabla _{j}\right) =\sum_{i=1}^{A}a_{\alpha
,i}a_{\alpha ,i}\nabla _{i}^{2}+\sum_{i,j=1\left( i\neq j\right)
}^{A}a_{\alpha ,i}a_{\alpha ,j}\left( \nabla _{i}\cdot \nabla _{j}\right) . 
\]

The mean value of this operator is simply 
\begin{eqnarray*}
\left\langle \nabla _{{\bf \eta }_{\alpha }}^{2}\right\rangle &\equiv
&\sum_{\sigma _{1,}\sigma _{2}...\sigma _{A}}\sum_{\tau _{1},\tau
_{2}...\tau _{A}}\int d{\bf r}_{1}\int d{\bf r}_{2}...\int d{\bf r}_{A}\Phi
_{{\cal E}J\Pi M_{T}}^{\ast }\left( {\bf r}_{1}{\bf r}_{2}...{\bf r}%
_{A}\sigma _{1,}\sigma _{2}...\sigma _{A}\tau _{1},\tau _{2}...\tau
_{A}\right) \\
&&\times \nabla _{{\bf \eta }_{\alpha }}^{2}\Phi _{{\cal E}J\Pi M_{T}}\left( 
{\bf r}_{1}{\bf r}_{2}...{\bf r}_{A}\sigma _{1,}\sigma _{2}...\sigma
_{A}\tau _{1},\tau _{2}...\tau _{A}\right) \\
&=&\sum_{i=1}^{A}a_{\alpha ,i}a_{\alpha ,i}\left\langle \nabla
_{i}^{2}\right\rangle +\sum_{i,j=1\left( i\neq j\right) }^{A}a_{\alpha
,i}a_{\alpha ,j}\left\langle \left( \nabla _{i}\cdot \nabla _{j}\right)
\right\rangle .
\end{eqnarray*}

Due to the antisymmetry of the wave functions, the matrix elements $%
\left\langle \nabla _{i}^{2}\right\rangle $ and $\left\langle \left( \nabla
_{i}\cdot \nabla _{j}\right) \right\rangle $ are the same for all possible
values of $i$ and $j$. Thus, 
\begin{eqnarray*}
\left\langle \nabla _{{\bf \eta }_{\alpha }}^{2}\right\rangle
&=&\left\langle \nabla _{A}^{2}\right\rangle \sum_{i=1}^{A}a_{\alpha
,i}a_{\alpha ,i}+\left\langle \left( \nabla _{A-1}\cdot \nabla _{A}\right)
\right\rangle \sum_{i,j=1\left( i\neq j\right) }^{A}a_{\alpha ,i}a_{\alpha
,j} \\
&=&\left\langle \nabla _{A}^{2}\right\rangle +\left\langle \left( \nabla
_{A-1}\cdot \nabla _{A}\right) \right\rangle \left[ \sum_{i=1}^{A}a_{\alpha
,i}\sum_{j=1}^{A}a_{\alpha ,j}-\sum_{i=1}^{A}a_{\alpha ,i}a_{\alpha ,i}%
\right] \\
&=&\left\langle \nabla _{A}^{2}\right\rangle -\left\langle \left( \nabla
_{A-1}\cdot \nabla _{A}\right) \right\rangle ,
\end{eqnarray*}
because all sums are independent of $\alpha $ , see Eqs. (\ref{3.1}) and (%
\ref{3.2}). Therefore, $\left\langle \nabla _{{\bf \eta }_{\alpha
}}^{2}\right\rangle $ is independent of $\alpha $ $\left( \alpha
=1,2,...,A-1\right) $.

\section{APPENDIX C}

The relation between potentials, defined in Eq. (\ref{20}), holds also for
the Reduced Hamiltonians (RH) of the different two-nucleon states 
\[
h^{0,j\pi t}=\sum_{m_{t}}c_{tm_{t}}h^{j\pi tm_{t}} 
\]
and their matrix elements 
\[
h_{\varepsilon ,\epsilon ^{\prime }}^{0,j\pi
t}=\sum_{m_{t}}c_{tm_{t}}h_{\varepsilon ,\epsilon ^{\prime }}^{j\pi tm_{t}}, 
\]
because the partial-wave expressions for the kinetic energy operators are
the same in all cases. For this reason, the difference of a diagonal matrix
element of the Hamiltonian $H\ $and an eigenvalue of the Hamiltonian $H^{0}$
for an arbitrary eigenstate of $H^{0}$ equals 
\begin{eqnarray}
\left\langle H-H^{0}\right\rangle _{{\cal E},{\cal E}}^{J\Pi TM_{T}} &=&%
\frac{A\left( A-1\right) }{2}  \nonumber \\
&&\times \sum_{\varepsilon ,\varepsilon ^{\prime },j\pi
tm_{t}}h_{\varepsilon ,\varepsilon ^{\prime }}^{j\pi tm_{t}}\left\{ 
\begin{array}{c}
Q_{\varepsilon ^{\prime },j\pi tm_{t};\varepsilon ,j\pi tm_{t}}\left( {\cal E%
}J\Pi TM_{T}\right) \\ 
-\delta _{\varepsilon ^{\prime },\varepsilon }c_{tm_{t}}Q_{\varepsilon ,j\pi
t}\left( {\cal E}J\Pi T\right)
\end{array}
\right\} .  \eqnum{C1}
\end{eqnarray}

There are nondiagonal contributions to the charge-dependent density matrix,
because the eigenfunctions of the three RH operators $h^{j\pi tm_{t}}$ ($%
m_{t}=0,\pm 1$) are different and are not equal to the eigenfunctions of the
isoscalar RH $h^{0,j\pi t}$. The matrix of the isoscalar RH in this case is
diagonal, i.e., 
\[
h_{\varepsilon ,\varepsilon ^{\prime }}^{0,j\pi
t}=\sum_{m_{t}}c_{tm_{t}}h_{\varepsilon ,\varepsilon ^{\prime }}^{j\pi
tm_{t}}=\varepsilon ^{j\pi t}\delta _{\varepsilon ,\varepsilon ^{\prime }}. 
\]

The left-hand side of Eq. (C1) is the sum of traces of products of two
symmetrical matrices, the first of these is the matrix of the RH and the
second is the reduced density matrix. As is seen from (C1), this sum
contains only two kinds of terms - one is equal to the product of diagonal
elements of both matrices, and the other is equal to the product of
nondiagonal elements. The last subsum is independent of the choice of the
values of variational coefficients $c_{tm_{t}}$. In any case, its
contribution is negligible in comparison with the first subsum, which
contains diagonal elements. This result is caused by the following reasons:
The nondiagonal elements of the charge-dependent RH matrices are negligible
in every two nucleon state, because charge-independence and charge-symmetry
breaking forces in light atomic nuclei are significantly weaker than
original charge-independent strong interaction between nucleons. The
nondiagonal entries of charge-dependent density matrix are also small in
absolute value, because eigenfunctions of the charge-dependent RH are close
to eigenfunctions of charge-independent RH for the same reason as given
previously.

The most important diagonal part of the sum in Eq. (C1) is 
\begin{equation}
\sum_{\varepsilon ,j\pi tm_{t}}h_{\varepsilon ,\varepsilon }^{j\pi
tm_{t}}\left\{ Q_{\varepsilon ,j\pi tm_{t}}\left( {\cal E}J\Pi TM_{T}\right)
-c_{tm_{t}}Q_{\varepsilon ,j\pi t}\left( {\cal E}J\Pi T\right) \right\} . 
\eqnum{C2}
\end{equation}
The problem is to choose the coefficients $c_{tm_{t}}$ in such a way that
this sum will have its minimal value. Let us first recall some definitions.
The distributions of probabilities, given in Eqs. (\ref{15}) and (\ref{18}),
differ. In the last expression they are comparable, because corresponding
density matrices are created by the same wave-function. The relationship\
between these probabilities, following directly from the definition of the
density matrix, is (see \cite{Kam81} and \cite{BBK83}): 
\begin{equation}
\sum_{m_{t}}Q_{\varepsilon ,j\pi tm_{t}}\left( {\cal E}J\Pi TM_{T}\right)
=Q_{\varepsilon ,j\pi t}\left( {\cal E}J\Pi T\right) .  \eqnum{C3}
\end{equation}
This equality means that the quantity on the right-hand side is a {\em %
marginal probability} (see, for example, \cite{Tikteor}). In the case $t=0$
, all entries of (C2) vanish, when $c_{00}=1$. This value is consistent with
the definition of the isoscalar interaction, Eqs. (\ref{20}), (\ref{20.0}),
because, as mentioned earlier, in this case we have no other choice,
excluding the best one. In the case when $t=1$, the situation is not so
simple, because only the sum of the three multipliers of $h_{\varepsilon
,\varepsilon }^{j\pi tm_{t}}$, corresponding to the three possible values of
the isospin projection $m_{t},$ vanishes, due to relations (\ref{20.0}) and
(C3). In general, for any value of $t$ the relation: 
\[
\sum_{m_{t}}\{Q_{\varepsilon ,j\pi tm_{t}}\left( {\cal E}J\Pi TM_{T}\right)
-c_{tm_{t}}Q_{\varepsilon ,j\pi t}\left( {\cal E}J\Pi T\right) \}=0 
\]
is satisfied.

This relation is an identity, and, hence, is not sufficient to define the
coefficients, but gives us more choices for determining the minimal value of
the sum given in Eq. (C2). The best choice occurs when the partial sum of
the coefficients over all possible two-nucleon channels is equal to zero,
namely, 
\[
\sum_{\varepsilon ,j\pi }\left\{ Q_{\varepsilon ,j\pi tm_{t}}\left( {\cal E}%
J\Pi TM_{T}\right) -c_{tm_{t}}Q_{\varepsilon ,j\pi t}\left( {\cal E}J\Pi
T\right) \right\} =0. 
\]
This equation can also be expressed in the form 
\begin{equation}
Q_{tm_{t}}\left( TM_{T}\right) =c_{tm_{t}}Q_{t}\left( T\right) .  \eqnum{C4}
\end{equation}
The marginal probabilities are defined by 
\[
\sum_{\varepsilon ,j\pi }Q_{\varepsilon ,j\pi tm_{t}}\left( {\cal E}J\Pi
TM_{T}\right) =Q_{tm_{t}}\left( TM_{T}\right) 
\]
and 
\[
\sum_{m_{t}}Q_{tm_{t}}\left( TM_{T}\right) =Q_{t}\left( T\right) 
\]
and can be extracted from the expressions for the simple multiparticle
operators with known eigenvalues, such as the mass number $A$, the charge $Z$
and the total isospin $T$, (see \cite{Kam81}, \cite{BBK83}). Therefore, 
\begin{eqnarray*}
Q_{00}\left( TM_{T}\right) &\equiv &Q_{0}\left( T\right) =\frac{A\left(
A+2\right) -4T\left( T+1\right) }{4A\left( A-1\right) }, \\
Q_{1}\left( T\right) &=&\frac{3A\left( A-2\right) +4T\left( T+1\right) }{%
4A\left( A-1\right) }, \\
Q_{1-1}\left( TM_{T}\right) &=&\frac{\left( A-2M_{T}\right) \left(
A-2M_{T}-2\right) }{4A\left( A-1\right) }, \\
Q_{10}\left( TM_{T}\right) &=&\frac{A\left( A-2\right) +4T\left( T+1\right)
-8M_{T}^{2}}{4A\left( A-1\right) }, \\
Q_{1+1}\left( TM_{T}\right) &=&\frac{\left( A+2M_{T}\right) \left(
A+2M_{T}-2\right) }{4A\left( A-1\right) }.
\end{eqnarray*}
Using Eq. (C4), we obtain 
\begin{equation}
c_{tm_{t}}=Q_{tm_{t}}\left( TM_{T}\right) /Q_{t}\left( T\right) .  \eqnum{C5}
\end{equation}
Applying this expression, we obtain the values of the coefficients given in
Eqs. (\ref{25})-(\ref{28}).

\bigskip

TABLE \ I. Comparison of energies, in MeV, obtained in the no-core
shell-model calculations with the full charge-dependent effective
interactions, E$_{P-N},$ and the corresponding isospin-invariant effective
interactions, E$_{ISO}$ (as defined in the text), for A=3 and A=6 nuclei.
The effective interactions were derived from the CD-Bonn potential \cite
{CDBonn}. Harmonic-oscillator energies of $\hbar \Omega $ =19 MeV and $\hbar
\Omega $ =17.2 MeV were employed for the A=3 and A=6 systems, respectively.
For A=3 a complete 8$\hbar \Omega $ model-space was utilized, while for A=6
a complete 6$\hbar \Omega $ model-space was used.

\begin{tabular}{llll}
\hline\hline
$^{A}$Z$\left( J^{\pi }T\right) $ \ \ \ \ \ \ \ \ \  & E$_{P-N}$\ \ \ \ \ \
\ \  & E$_{ISO}$ \ \ \ \ \ \ \ \ \  & $\left[ E_{P-N}-E_{ISO}\right]
/E_{P-N} $ \\ \hline
$^{3}$H$\left( \frac{1}{2}^{+}\frac{1}{2}\right) $ & -8.441 & -8.421 & $%
-2.44\times 10^{-3}$ \\ 
$^{3}$He$\left( \frac{1}{2}^{+}\frac{1}{2}\right) $ & -7.668 & -7.643 & $%
-3.27\times 10^{-3}$ \\ 
$^{6}$He$\left( 0^{+}1\right) $ & -25.665 & -25.830 & $+6.42\times 10^{-3}$
\\ 
$^{6}$He$\left( 2^{+}1\right) $ & -22.944 & -23.085 & $+6.13\times 10^{-3}$
\\ 
$^{6}$Li$\left( 1^{+}0\right) $ & -28.257 & -28.180 & $-2.73\times 10^{-3}$
\\ 
$^{6}$Li$\left( 3^{+}0\right) $ & -25.605 & -25.525 & $-3.12\times 10^{-3}$
\\ 
$^{6}$Li$\left( 0^{+}1\right) $ & -24.815 & -24.666 & $-6.02\times 10^{-3}$
\\ 
$^{6}$Li$\left( 2^{+}0\right) $ & -23.146 & -23.071 & $-3.25\times 10^{-3}$
\\ 
$^{6}$Li$\left( 2^{+}1\right) $ & -21.934 & -21.924 & $-4.47\,\times 10^{-4}$
\\ 
$^{6}$Li$\left( 1^{+}0\right) $ & -20.433 & -20.362 & $-3.46\times 10^{-3}$
\\ 
$^{6}$Be$\left( 0^{+}1\right) $ & -22.829 & -22.545 & $-1.25\times 10^{-2}$
\\ 
$^{6}$Be$\left( 2^{+}1\right) $ & -20.227 & -19.851 & $-1.86\times 10^{-2}$
\\ \hline\hline
\end{tabular}

\newpage

\subsection{Figure captions}

FIG. 1. An example of a Jacobi tree.

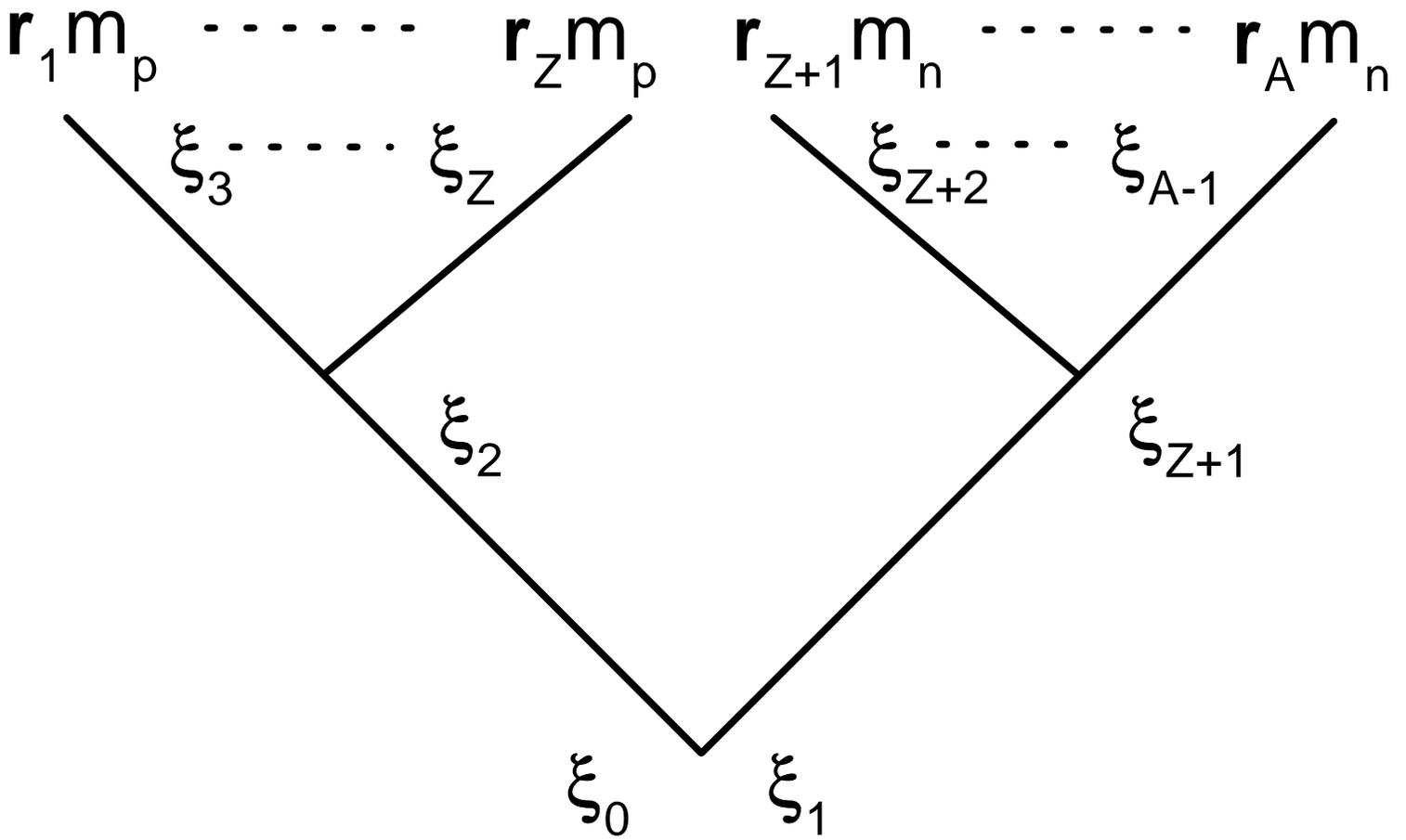
FIG. 2. The Jacobi tree for a nucleus composed of Z protons and N neutrons.

FIG. 3. An example of a Jacobi tree with subtrees.

\end{document}